# Crystal growth and optical properties of Ce-doped (La,Y)$_2$Si$_2$O$_7$ single crystal

Takahiko Horiai[1,*], Juraj Paterek[1], Jan Pejchal[1], Marketa Jarosova[1], Jan Rohlicek[1], Shunsuke Kurosawa[2,3], Takashi Hanada[2], Masao Yoshino[2], Akihiro Yamaji[2,3], Satoshi Toyoda[2,3], Hiroki Sato[2,3], Yuji Ohashi[2,3], Kei Kamada[3,4], Yuui Yokota[2], Akira Yoshikawa[2,3,4], and Martin Nikl[1]

[1] Institute of Physics, Czech Academy of Sciences, Cukrovarnicka 10, Prague, Czech Republic
[2] Institute for Materials Research, Tohoku University, 2-1-1 Katahira, Sendai, Japan
[3] New Industry Creation Hatchery Center, Tohoku University, 6-6-10 Aoba, Sendai, Japan
[4] C&A corporation, 1-16-23 Ichibancho, Sendai, Japan
*E-mail: horiai@fzu.cz

## Abstract

We have grown Ce-doped (La,Y)$_2$Si$_2$O$_7$ single crystal by micro-pulling-down method and investigated its optical and scintillation properties. We have successfully prepared the single crystal with (Ce$_{0.015}$La$_{0.600}$Y$_{0.385}$)$_2$Si$_2$O$_7$ composition. The observed thermal quenching process could be characterized by the quenching temperature ($T_{50\%}$) of 526 K and its activation energy was determined to be 0.62 eV. Further considering the thermal quenching factors, it was found that the thermal quenching was caused by at least the thermal ionization and maybe also by classical thermal quenching. The light output and scintillation decay time were evaluated to be ~12,000 photons/MeV and ~42 ns, respectively. These results indicate that (Ce$_{0.015}$La$_{0.600}$Y$_{0.385}$)$_2$Si$_2$O$_7$ has a great potential for application in scintillation materials.

**Introduction**

Scintillation crystals convert energy of ionizing radiation to ultra-violet or visible light, and are widely used in radiation detector applications such as oil well logging, medical imaging and high energy physics [1-5]. There are two types of scintillation crystals: organic scintillators and inorganic scintillators. Inorganic scintillators have higher light output, density and atomic number than organic scintillators, which makes them more suitable for detection of gamma-rays and X-rays. In particular, the optical and the scintillation characteristics of Ce-doped oxide scintillators, e.g. orthosilicates ($Gd_2SiO_5$), garnets ($Gd_3(Al,Ga)_5O_{12}$) and perovskites ($YAlO_3$), have been investigated [6–9].

Recently, Ce-doped pyrosilicate scintillators ($RE_2Si_2O_7$, RE=rare-earth ion), e.g. $Lu_2Si_2O_7$ and $(La,Gd)_2Si_2O_7$ (Ce:La-GPS), have attracted much attention due to their high light output, relatively short decay time and good thermal stability of the light output [10–14]. The optical characteristics and temperature dependence of nanoseconds decay times of $(Gd_{0.52}La_{0.48})_2Si_2O_7$:Ce1% were reported by Jary et al. and demonstrated that the thermal quenching occurs above about 440 K [15]. As a result, it is expected that pyrosilicate scintillators, which have a good thermal stability, are promising materials for application in oil well logging, since the soil temperature reaches up to 500 K g in the shale oil layer.

However, most of the compositions such as $RE_2Si_2O_7$ (RE=Y, La, Gd, etc.) are not congruent in the $RE_2O_3$-$SiO_2$ systems which makes them difficult to be prepared as single crystals from the melt [16]. A recent report on the single crystal growth of $Gd_2Si_2O_7$ demonstrated that the substitution of $Gd^{3+}$ by $Ce^{3+}$ increases the average ionic radius of the Gd site and stabilizes the pyrosilicate phase [17]. However, the self-absorption and the concentration quenching occurred due to high Ce-concentration. In order to prevent the concentration quenching, Suzuki et al. stabilized the pyrosilicate phase by substituting $Gd^{3+}$ by $La^{3+}$, which has similar ionic radius as $Ce^{3+}$, and succeeded in growing a single crystal using floating zone method [12]. Furthermore, the single crystal growth of Ce:La-GPS with various $Gd^{3+}$:$La^{3+}$ ratios by micro-pulling-down (μ-PD) method and Czochralski technique have been reported [12,14,18,19].

In the development of other RE pyrosilicate scintillators, considering the phase diagram of $La_2Si_2O_7$-$Y_2Si_2O_7$ pseudo-binary system reported by Toropov and Mandal, an invariant point corresponding to the $(La_{0.6}Y_{0.4})_2Si_2O_7$ composition was found [20]. In addition, Fernández-Carrión et al. found that the $La_2Si_2O_7$-$Y_2Si_2O_7$ system shows a solid solubility region of the monoclinic system with the space group $P2_1/c$ in 0-50% $Y_2Si_2O_7$ range above 1300°C [21]. Considering these previous studies, we expect to be able to grow single crystal of $(La_{0.6}Y_{0.4})_2Si_2O_7$. Thus, in this study, we tried to grow Ce-doped $(La,Y)_2Si_2O_7$ single crystal and investigated its optical and scintillation properties and thermal stability.

**Materials and methods**

The $(Ce_{0.015}La_{0.600}Y_{0.385})_2Si_2O_7$ (Ce:La60-YPS) single crystal was grown from the melt using the µ-PD method. As starting materials, we used $Y_2O_3$, $La_2O_3$, $CeO_2$ and $SiO_2$ powders with a purity of more than 99.99%. After weighing and mixing the powders in the ratio corresponding to the nominal composition of Ce:La60-YPS, crystal growth was performed using Ir crucible and Ce:La-GPS as a seed crystal. In order to prevent oxidation of the Ir crucible, $N_2$ gas flowed through the hot-zone during the crystal growth. The pulling down rate was 0.01~0.03 mm/min and the post growth cooling time was 4~6 hours.

After the crystal growth, the chemical composition in the grown crystal was determined by the electron probe microanalyzer (EPMA, JXA-8230, JEOL) equipped with wavelength dispersive spectrometers (WDS). In addition, the powder X-ray diffraction (XRD) analysis (SmartLab, Rigaku) was performed to verify the crystal phase with an X-ray generator with the accelerating voltage of 45 kV and the beam current of 200 mA using Cu-target. The XRD patterns were measured with Cu Kα radiation with 0.01º step in the 5-100º 2-theta range.

To evaluate the optical characteristics of Ce:La60-YPS, the photoluminescence (PL) excitation and emission spectra were measured with two monochromators (SF-4 and SPM-1). The laser-driven light source (EQ-99X, Energetiq) was used as the excitation source. Furthermore, the PL decay curves were measured in the 150-700 K temperature range, and were evaluated for thermal quenching. The 339 nm nanosecond light emitting diodes (nanoLED) was used as an excitation source, and the photomultiplier tube (PMT, TBX-04, IBH Scotland) was used for detection by the time-correlated single photon counting technique. The decay profiles were deconvoluted using two exponential approximation to estimate exact decay times.

The scintillation light output was evaluated under gamma-rays excitation ($^{137}$Cs, 662 keV) using the PMT (R7600U-200, Hamamatsu), where the sample was covered with Teflon$^{TM}$ tape as a reflector. The PMT signals were amplified (Model 113, ORTEC), shaped with shaping time of 2 µs (Model 752A, ORTEC), and read out with the pocket multi-channel analyzer (8000A, Amptek) to the personal computer. The decay curves excited by gamma-rays from a $^{137}$Cs source were measured with the oscilloscope (TDS3052B, Tektronix), and were fitted by a function consisting of two exponential components.

**Results and discussion**

$(Ce_{0.015}La_{0.600}Y_{0.385})_2Si_2O_7$ was grown by the µ-PD method and the grown crystal is shown in Fig 1. The grown crystal was 3.5 mm in diameter and 25 mm in length. The beginning and the end parts of the grown crystal were not transparent, however we succeeded in growing a transparent crystal in the central part. The grown crystal was cut and mirror polished to a thickness of 1 mm for evaluation of the optical and the scintillation properties. In order to identify the exact chemical composition of the grown

Ce:La60-YPS, the WDS-EPMA measurement was performed. The result of the composition analysis by EPMA is listed in Table 1. The exact chemical composition in the grown crystal was consistent with the nominal composition. In addition, the powder XRD measurement was performed to identify the crystalline system. Figure 2 shows the diffraction patterns of Ce:La60-YPS in the 5-100° 2-theta range. From this diffraction patterns, the crystalline system of Ce:La60-YPS was identified to be monoclinic $P2_1/c$, and it was consistent with the previous reports (JCPDS No. 53-0945). Considering these results, we revealed that it is possible to grow $(La,Y)_2Si_2O_7$ single crystal at specific La:Y ratio.

After cutting and mirror polishing the sample to the thickness of 1 mm, the PL excitation and emission spectra at 300 K were measured. As shown in Fig. 3, the emission peaks associated with the $Ce^{3+}$ $5d_1$-$^2F_{5/2}$ and $5d_1$-$^2F_{7/2}$ transitions were observed at 368 nm and 385 nm, respectively. Regarding the excitation spectrum, the excitation peaks related to $^2F_{5/2}$-$5d_{1,2,3,4,5}$ were observed around 343 nm, 323 nm, 275 nm, 248 nm and 219 nm, respectively. In addition, the temperature dependence of the PL emission spectra was evaluated in 100-700 K temperature range (Fig. 4). The PL intensity gradually decreased with increasing temperature. It is worth noting that the emission intensity of $Ce^{3+}$ $5d_1$-$^2F_{5/2}$ transition is larger than that of $Ce^{3+}$ $5d_1$-$^2F_{7/2}$ transition at low temperatures, while the emission intensity of $Ce^{3+}$ $5d_1$-$^2F_{7/2}$ transition becomes larger than that of $Ce^{3+}$ $5d_1$-$^2F_{5/2}$ transition as the temperature increases. Considering the fact that the emission peak of $Ce^{3+}$ $5d_1$-$^2F_{5/2}$ transition shifts to longer wavelength with increasing temperature and that the peak emission of $Ce^{3+}$ $5d_1$-$^2F_{7/2}$ transition remains constant, the thermally-enhanced re-absorption of $Ce^{3+}$ ion luminescence is expected to take place.

For further evaluation of the thermal quenching of Ce:La60-YPS, the PL decay curves under excitation by 339 nm nanoLED were measured at different temperatures (Figure 5). These decay curves were fitted by two exponential functions applying the instrumental response function correction. Figure 6 shows the PL decay times of Ce:La60-YPS as a function of the temperature. In the temperature of 150-400 K, the PL decay times were kept constant at approximately 27 ns. Above 400 K, the PL decay time decreased, and the temperature quenching occurred. To evaluate the activation energy ($\Delta E$) of the thermal quenching and the thermal quenching temperature ($T_{50\%}$), the fitting was performed with a single-barrier model shown in the following equation:

$$\tau(T) = \frac{1}{\Gamma_r + \Gamma_n \exp(-\Delta E/k_B T)}$$

where $\tau$ is the PL decay time, $\Gamma_r$ is the radiative rate, $\Gamma_n$ is the attempt rate of nonradiative process, $k_B$ is the Boltzmann constant (8.6171×10$^{-5}$ [eV K$^{-1}$]) and $T$ is the temperature. As a result of fitting, $\Gamma_n$ and $\Delta E$ were found to be 2.85×10$^{13}$ s$^{-1}$ and 0.62 eV, respectively. Moreover, the $T_{50\%}$, defined as the temperature at which the PL decay time becomes 50% of the low temperature value, was determined to be 526 K. The $\Delta E$ and $T_{50\%}$ values of the Ce:La60-YPS were relatively high and comparable to the results of Ce:La-GPS [15].

Generally, the thermal quenching is caused by several factors: (1) thermal ionization induced by the energy transition of electrons from the $Ce^{3+}$ 5d levels to the conduction band, (2) thermally activated nonradiative crossover from the $Ce^{3+}$ $5d_1$ excited level to the 4f ground state and (3) simple or thermally activated concentration quenching by nonradiative energy migration among $Ce^{3+}$ ions to killer centers. Here in Fig. 6, the temperature dependence of the background intensity of the PL decay time is also indicated, and its increase was observed at above 400 K. Taking the results of the temperature dependence of PL emission spectra and PL decay time into account, the thermal quenching in Ce:La60-YPS is considered to be caused by at least the thermal ionization. Classical thermal quenching due to thermally activated nonradiative crossover from the $Ce^{3+}$ $5d_1$ excited level to the 4f ground state can play role as well.

To evaluate the light output of Ce:La60-YPS, the pulse height spectrum was measured. Figure 7 shows the pulse height spectra of Ce:La60-YPS and $Ce:Gd_2SiO_5$ used as reference under excitation with $^{137}Cs$ gamma-rays source. The photoelectric absorption peak for 662 keV gamma-rays of Ce:La60-YPS was observed at about channel 170. The light output of Ce:La60-YPS was determined to be ~12,000 photons/MeV by comparison with the photoelectric absorption peak channel of $Ce:Gd_2SiO_5$ (light output: 11,000 photons/MeV). The light output of Ce:La60-YPS was lower than that of Ce:La-GPS (~41,000 photons/MeV [22]), which is most probably due to the fact that the efficient energy transfer from $Gd^{3+}$ to $Ce^{3+}$ ions occurs in Ce:La-GPS. Figure 8 illustrates the scintillation decay curve of Ce:La60-YPS excited by gamma-rays using a $^{137}Cs$ source. To determine the scintillation decay time, the decay curve was fitted by two exponential function. The scintillation decay times were determined to be ~42 ns (48.3%) for the fast component and ~311 ns (51.7%) for the slow component, respectively.

**Conclusions**

The transparent $(Ce_{0.015}La_{0.600}Y_{0.385})_2Si_2O_7$ single crystal was successfully grown by the μ-PD method. The crystalline system was identified to be monoclinic (space group: $P2_1/c$). Regarding the optical characteristics, the typical broad $5d_1$-4f emission of $Ce^{3+}$ ions was observed, and the thermally-enhanced re-absorption was observed. The quenching temperature and the activation energy were determined to be 526 K and 0.62 eV, respectively, by fitting the results of the temperature dependence of the PL decay time using a single-barrier model. In addition, we found that the thermal quenching of Ce:La60-YPS was caused by at least the thermal ionization and maybe also by classical thermal quenching. Using $^{137}Cs$ gamma-rays source, we evaluated the light output and the scintillation decay time, which were determined to be ~12,000 photons/MeV and ~42 ns, respectively. These scintillation properties were comparable to those of the Ce-doped $Gd_2SiO_5$, which is widely applied, and furthermore, Ce:La60-YPS shows the good thermal stability, so it is suitable for applications such as the oil well logging.

**Acknowledgments**

This work is partially supported by (1) Japan Society for the Promotion of Science (JSPS) KAKENHI Grant Number 15597934(15K13478), 15619740(15K18209), (17H05190), (19H02422), (19H04684), 16818695(16H06633), (17J03074) (by Grant-in-Aid for Young Scientists(B), etc., S. Kurosawa, A. Yamaji, T. Horiai), (2) Leading Initiative for Excellent Young Researchers (LEADER), MeXT, Grant Number 16809648, (3) The Czech Academy of Sciences and JSPS under the Japan - Czech Republic Research Cooperative Program, (4) Japan Agency for Medical Research and Development (AMED), Medical Research and Development Programs Focused on Technology Transfers: Development of Advanced Measurement and Analysis Systems (AMED-SENTAN), (5) Japan Science and Technology Agency (JST), Competitive Funding Programs A-STEP (Adaptable & Seamless Technology Transfer Program through Target-driven R&D), Grant Number JPMJTS1521, (6) New Energy and Industrial Technology Development Organization (NEDO), (7) the Nuclear Energy Science & Technology and Human Resource Development Project (through concentrating wisdom) from the Ministry of Education, Culture, Sports, Science and Technology of Japan(MeXT), (8) The Murata Science Foundation, (9) Yazaki Memorial Foundation for Science and Technology, (10) Tokin Science and Technology Promotion Foundation, (11) Intelligent cosmos research institute, (12) Terumo Life Science Foundation, Special research subsidy, (13) Iketani Science and Technology Foundation, (14) The Sumitomo Foundation, grant for Basic Science research projects, (15) Frontier Research Institute for Interdisciplinary Sciences, Tohoku University, (16) International Collaboration Center Institute for Materials Research (ICC-IMR), Tohoku University, (17) Foundation for Promotion of Material Science and Technology of Japan, (18) UVSOR facility, Institute for Molecular Science (project number 19-527), (19) Czech Operational Programme Research, Development and Education financed by the European Structural and Investment Funds and the Czech Ministry of Education, Youth and Sports, project no. CZ.02.1.01/0.0/0.0/16_019/0000760 (SOLID21) and (20) Czech Operational Programme Research, Development and Education financed by the European Structural and Investment Funds and the Czech Ministry of Education, Youth and Sports, project no. CZ.02.2.69/0.0/0.0/18_053/0016627 ('FZU researchers, technical and administrative staff mobility'). In addition, we would like to thank following persons for their support: Mr. Yoshihiro Nakamura of Institute of Multidisciplinary Research for Advanced Materials (IMRAM), Tohoku University and Ai Kaminaka of NICHe, Tohoku University.

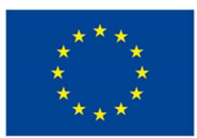
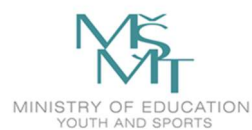

**Figure caption**

Figure 1. $(Ce_{0.015}La_{0.600}Y_{0.385})_2Si_2O_7$ crystal grown by μ–PD method.

Figure 2. XRD patterns of $(Ce_{0.015}La_{0.600}Y_{0.385})_2Si_2O_7$.

Figure 3. PL excitation and emission spectra of $(Ce_{0.015}La_{0.600}Y_{0.385})_2Si_2O_7$ at 300 K.

Figure 4. Temperature dependence of PL emission spectrum of $(Ce_{0.015}La_{0.600}Y_{0.385})_2Si_2O_7$.

Figure 5. PL decay curves of $(Ce_{0.015}La_{0.600}Y_{0.385})_2Si_2O_7$ excited by 339 nm nanoLED in the 150-700 K temperature range.

Figure 6. Temperature dependence of PL decay time and background intensity of $(Ce_{0.015}La_{0.600}Y_{0.385})_2Si_2O_7$.

Figure 7. Pulse height spectra of $(Ce_{0.015}La_{0.600}Y_{0.385})_2Si_2O_7$ and $Ce:Gd_2SiO_5$ used as reference with $^{137}Cs$ gamma-rays source.

Figure 8. Scintillation decay curve of $(Ce_{0.015}La_{0.600}Y_{0.385})_2Si_2O_7$ excited by gamma-rays using a $^{137}Cs$ source.

Table 1. Result of composition analysis by the EPMA for $(Ce_{0.015}La_{0.600}Y_{0.385})_2Si_2O_7$ [mass%].

|  | $Y_2O_3$ | $La_2O_3$ | $Ce_2O_3$ | $SiO_2$ |
|---|---|---|---|---|
| Nominal concentration ($C_0$) | 21.33 | 47.97 | 1.21 | 29.49 |
| Actual concentration ($C$) | 22.02 | 48.06 | 1.28 | 28.63 |
| $C / C_0$ | 1.03 | 1.00 | 1.06 | 0.97 |

**Figures**

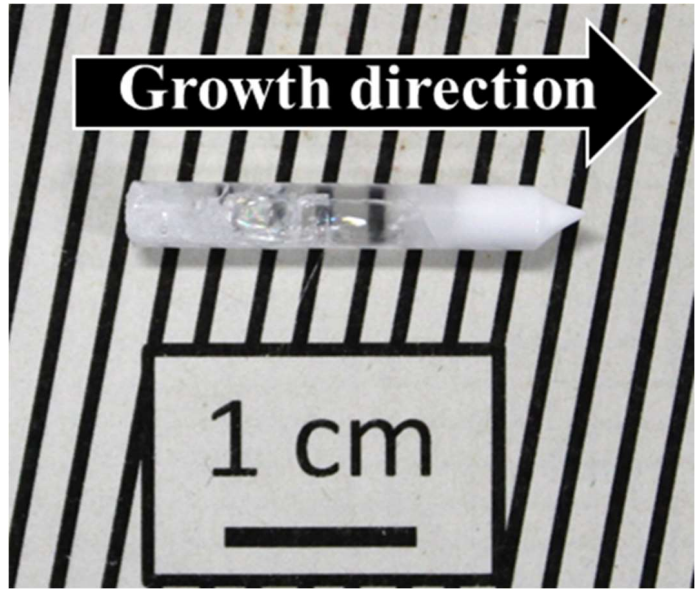

Fig. 1　$(Ce_{0.015}La_{0.600}Y_{0.385})_2Si_2O_7$ crystal grown by μ–PD method.

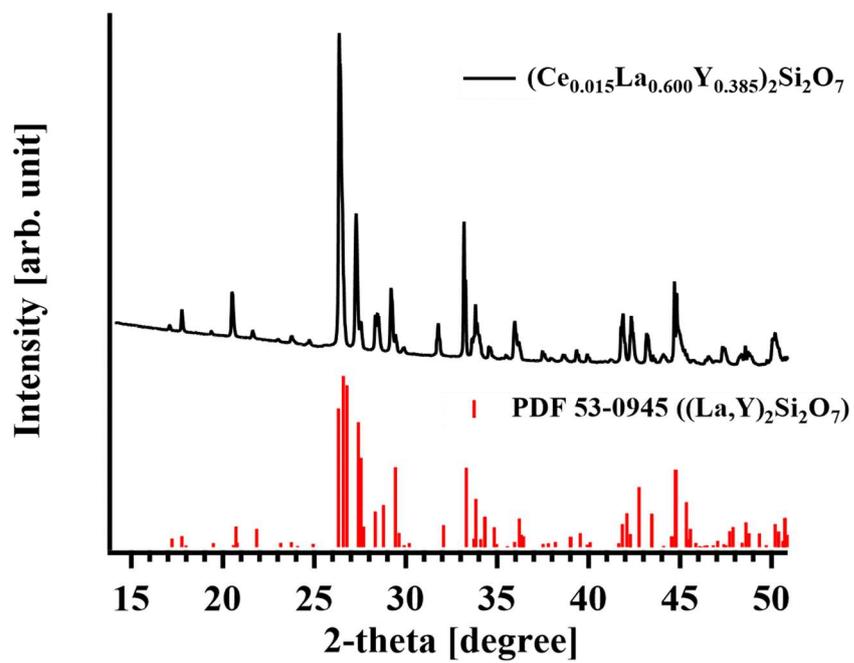

Fig. 2　XRD patterns of $(Ce_{0.015}La_{0.600}Y_{0.385})_2Si_2O_7$.

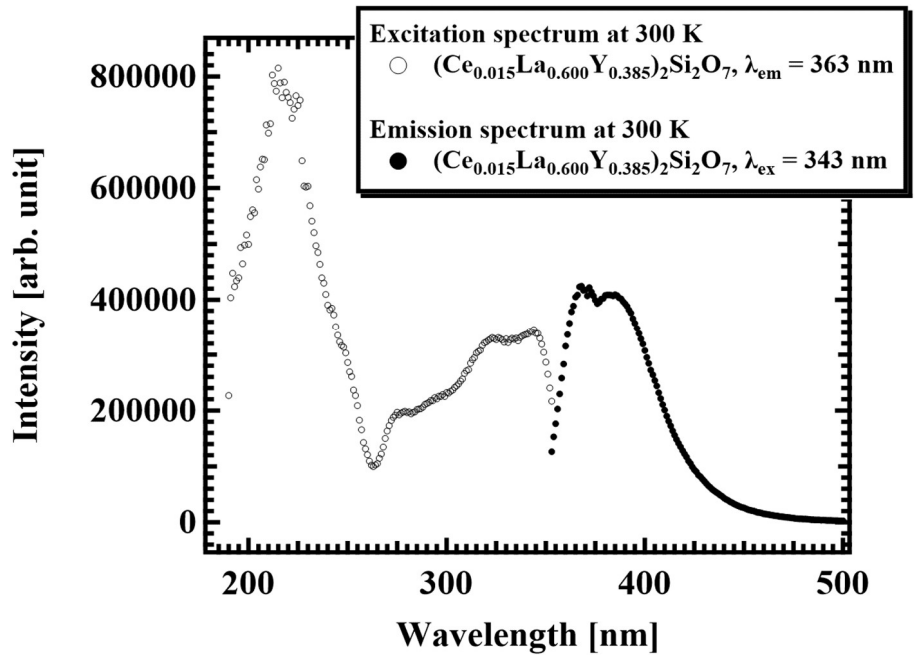

Fig. 3  PL excitation and emission spectra of $(Ce_{0.015}La_{0.600}Y_{0.385})_2Si_2O_7$ at 300 K.

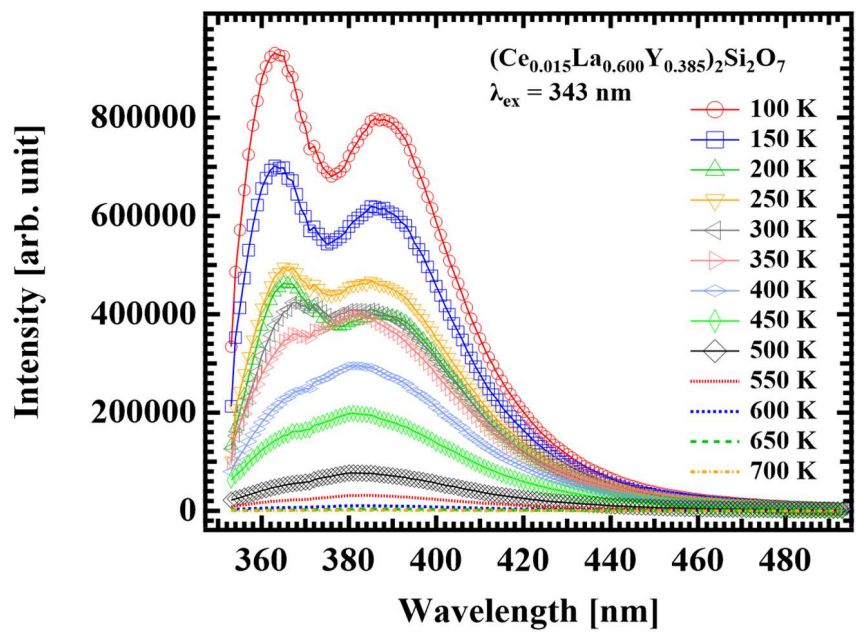

Fig. 4  Temperature dependence of PL emission spectrum of $(Ce_{0.015}La_{0.600}Y_{0.385})_2Si_2O_7$.

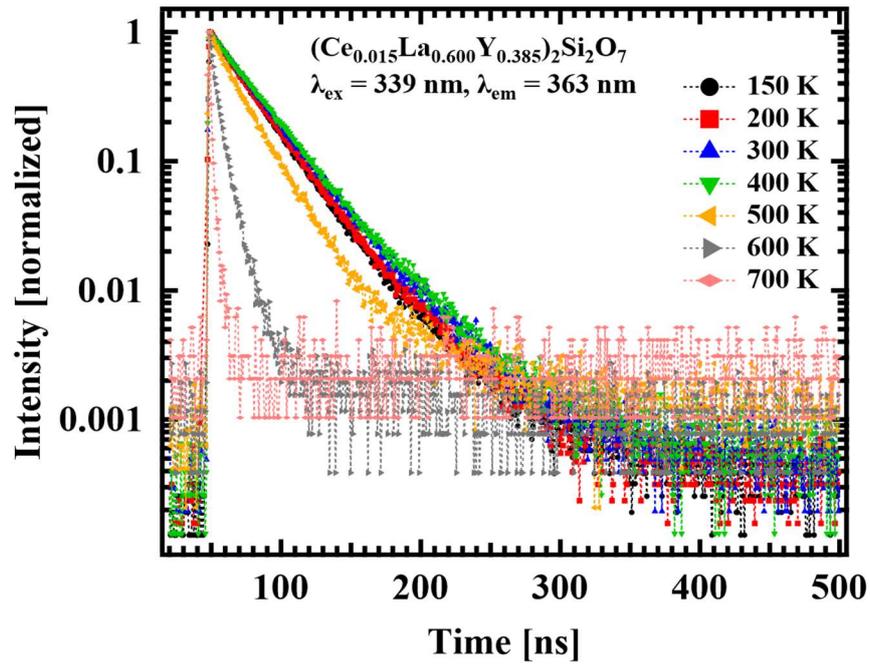

Fig. 5  PL decay curves of $(Ce_{0.015}La_{0.600}Y_{0.385})_2Si_2O_7$ excited by 339 nm nanoLED in the 150-700 K temperature range.

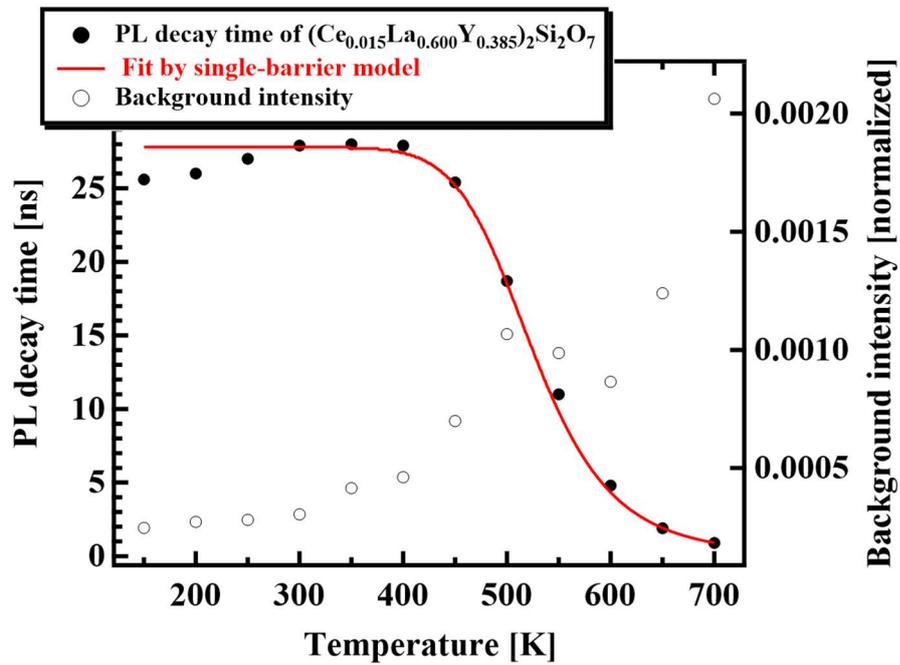

Fig. 6  Temperature dependence of PL decay time and background intensity of $(Ce_{0.015}La_{0.600}Y_{0.385})_2Si_2O_7$.

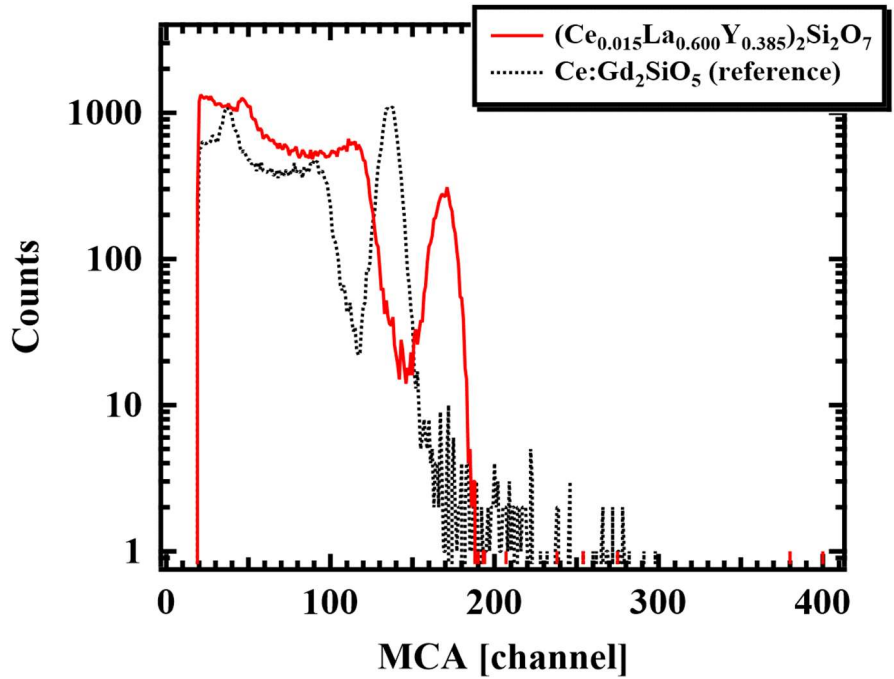

Fig. 7　Pulse height spectra of $(Ce_{0.015}La_{0.600}Y_{0.385})_2Si_2O_7$ and $Ce:Gd_2SiO_5$ used as reference with $^{137}Cs$ gamma-rays source.

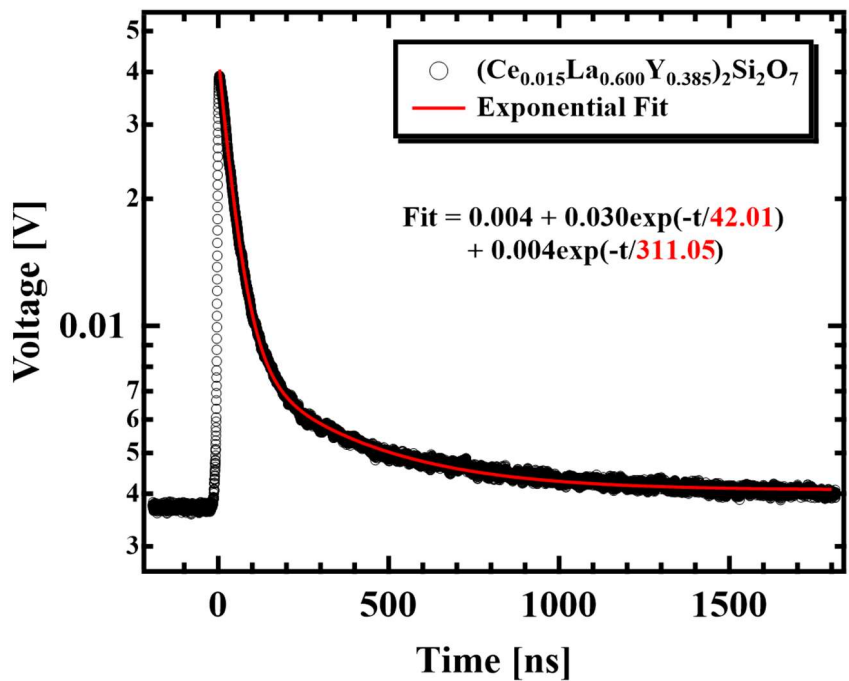

Fig. 8　Scintillation decay curve of $(Ce_{0.015}La_{0.600}Y_{0.385})_2Si_2O_7$ excited by gamma-rays using a $^{137}Cs$ source.